\documentclass[%
 aip,
 amsmath,amssymb,
 reprint,%
]{revtex4-1}

\usepackage{graphicx}
\usepackage{dcolumn}
\usepackage{bm}

\usepackage[utf8]{inputenc}
\usepackage[T1]{fontenc}
\usepackage{mathptmx}
\usepackage{etoolbox}
\usepackage{makecell}
\usepackage{xcolor}

\makeatletter
\def\@email#1#2{%
 \endgroup
 \patchcmd{\titleblock@produce}
  {\frontmatter@RRAPformat}
  {\frontmatter@RRAPformat{\produce@RRAP{*#1\href{mailto:#2}{#2}}}\frontmatter@RRAPformat}
  {}{}
}%
\makeatother
\begin{document}

\preprint{AIP/123-QED}
\title{Programmable entangled qubit states on a linear-optical platform}

\author{N.N. Skryabin}
\altaffiliation{These authors contributed equally to this work}
\email{nikolay.skryabin@phystech.edu}
\affiliation{%
 Quantum Technology Centre and Faculty of Physics, M.V. Lomonosov Moscow State University, 1 Leninskie Gory Street, Moscow 119991, Russia}%

\author{Yu.A. Biriukov}
\altaffiliation{These authors contributed equally to this work}
\affiliation{%
 Quantum Technology Centre and Faculty of Physics, M.V. Lomonosov Moscow State University, 1 Leninskie Gory Street, Moscow 119991, Russia}%

\author{M.A. Dryazgov}
\affiliation{%
 Quantum Technology Centre and Faculty of Physics, M.V. Lomonosov Moscow State University, 1 Leninskie Gory Street, Moscow 119991, Russia}%

\author{S.A. Fldzhyan}
\affiliation{%
 Quantum Technology Centre and Faculty of Physics, M.V. Lomonosov Moscow State University, 1 Leninskie Gory Street, Moscow 119991, Russia}%
\affiliation{%
 Russian Quantum Center, 30 Bolshoy bul'var building 1, Moscow 121205, Russia}%
  
\author{S.A. Zhuravitskii}
\affiliation{%
 Quantum Technology Centre and Faculty of Physics, M.V. Lomonosov Moscow State University, 1 Leninskie Gory Street, Moscow 119991, Russia}%

\author{A.S. Argenchiev}
\affiliation{%
 Quantum Technology Centre and Faculty of Physics, M.V. Lomonosov Moscow State University, 1 Leninskie Gory Street, Moscow 119991, Russia}%

\author{I.V. Kondratyev}
\affiliation{%
 Quantum Technology Centre and Faculty of Physics, M.V. Lomonosov Moscow State University, 1 Leninskie Gory Street, Moscow 119991, Russia}%

 \author{L.A. Tsoma}
\affiliation{%
 Quantum Technology Centre and Faculty of Physics, M.V. Lomonosov Moscow State University, 1 Leninskie Gory Street, Moscow 119991, Russia}%

 \author{K.I. Okhlopkov}
\affiliation{%
 Quantum Technology Centre and Faculty of Physics, M.V. Lomonosov Moscow State University, 1 Leninskie Gory Street, Moscow 119991, Russia}%
\affiliation{%
 Russian Quantum Center, 30 Bolshoy bul'var building 1, Moscow 121205, Russia}%

 \author{I.M. Gruzinov}
\affiliation{%
 Russian Quantum Center, 30 Bolshoy bul'var building 1, Moscow 121205, Russia}%

\author{K.V. Taratorin}
\affiliation{%
 Quantum Technology Centre and Faculty of Physics, M.V. Lomonosov Moscow State University, 1 Leninskie Gory Street, Moscow 119991, Russia}%

\author{M.Yu. Saygin}
\affiliation{%
 Quantum Technology Centre and Faculty of Physics, M.V. Lomonosov Moscow State University, 1 Leninskie Gory Street, Moscow 119991, Russia}%
\affiliation{Laboratory of Quantum Engineering of Light, South Ural State University (SUSU), 76 Prospekt Lenina, Chelyabinsk 454080, Russia}%
 
\author{I.V. Dyakonov}
\affiliation{%
 Quantum Technology Centre and Faculty of Physics, M.V. Lomonosov Moscow State University, 1 Leninskie Gory Street, Moscow 119991, Russia}%
\affiliation{%
 Russian Quantum Center, 30 Bolshoy bul'var building 1, Moscow 121205, Russia}%

\author{M.V. Rakhlin}
\affiliation{Ioffe Institute, St. Petersburg, 194021, Russia}

\author{A.I. Galimov}
\affiliation{Ioffe Institute, St. Petersburg, 194021, Russia}

\author{G.V. Klimko}
\affiliation{Ioffe Institute, St. Petersburg, 194021, Russia}

\author{S.V. Sorokin}
\affiliation{Ioffe Institute, St. Petersburg, 194021, Russia}

\author{I.V. Sedova}
\affiliation{Ioffe Institute, St. Petersburg, 194021, Russia}

\author{M.M. Kulagina}
\affiliation{Ioffe Institute, St. Petersburg, 194021, Russia}

\author{Yu.M. Zadiranov}
\affiliation{Ioffe Institute, St. Petersburg, 194021, Russia}

\author{A.A. Toropov}
\affiliation{Ioffe Institute, St. Petersburg, 194021, Russia}

\author{S.A. Evlashin}
\affiliation{%
Skolkovo Institute of Science and Technology, 30 Bolshoy bul'var building 1, Moscow 121205, Russia}%

\author{A.A. Korneev}
\affiliation{%
 Quantum Technology Centre and Faculty of Physics, M.V. Lomonosov Moscow State University, 1 Leninskie Gory Street, Moscow 119991, Russia}%

\author{S.P. Kulik}
\affiliation{%
 Quantum Technology Centre and Faculty of Physics, M.V. Lomonosov Moscow State University, 1 Leninskie Gory Street, Moscow 119991, Russia}%
\affiliation{Laboratory of Quantum Engineering of Light, South Ural State University (SUSU), 76 Prospekt Lenina, Chelyabinsk 454080, Russia}%

\author{S.S. Straupe}
\affiliation{%
 Quantum Technology Centre and Faculty of Physics, M.V. Lomonosov Moscow State University, 1 Leninskie Gory Street, Moscow 119991, Russia}%
\affiliation{%
 Russian Quantum Center, 30 Bolshoy bul'var building 1, Moscow 121205, Russia}%

\date{\today}

\begin{abstract}

We present an experimental platform for linear-optical quantum information processing. Our setup utilizes multiphoton generation using a high-quality single-photon source, which is demultiplexed across multiple spatial channels, a custom-designed, programmable, low-loss photonic chip, and paired with high-efficiency single-photon detectors. We demonstrate the platform's capability in producing heralded arbitrary two-qubit dual-rail encoded states, a crucial building block for large-scale photonic quantum computers. The programmable chip was fully characterized through a calibration process that allowed us to create a numerical model accounting for fabrication imperfections and measurement errors. As a result, using on-chip quantum state tomography (QST), we achieved high-fidelity quantum state preparation, with a fidelity of 98.5\% specifically for the Bell state. 

\end{abstract}

\maketitle

\section{\label{sec:Intro}Introduction}

The photonic platform of quantum computing is one of a few platforms that reached the quantum advantage threshold \cite{Zhong2020, Zhu2022, Madsen2022}. This platform offers several unique advantages, such as negligible decoherence even at room temperatures, accessibility of several convenient degrees of freedom to manipulate quantum states \cite{Flamini2019}, compatibility with fiber-optic communication networks. Moreover, it is pertinent to consider the rapidly advancing field of integrated quantum photonics \cite{Meany2015, Elshaari2020, Saravi2021}, including those based on CMOS production lines \cite{Wang2020, Adcock2021}. The most efficient approach to photonic quantum computing is linear-optical quantum computing \cite{Knill2001, Kok2007, OBrien2007}, which enables the generation of entangled states from individual single photons by feeding them to a network of linear optical elements only. However, the entanglement generation in linear circuits is not a deterministic process, rather, it occurs with a certain success probability \cite{Ralph2002}. This limitation makes it challenging to scale linear optical systems further, as the success probability decreases rapidly.

The measurement-based quantum computing (MBQC) \cite{Briegel2009} is a framework that maps a conventional quantum circuit onto a sequence of single-qubit measurements applied to an entangled cluster state that can be generated beforehand. This transition from coherent single- and two-qubit operations to a measurement-only algorithm is particularly well-suited for the photonic platform, as it eliminates the need to store photons throughout the entire duration of the algorithm. The generation of cluster states can be decomposed into generating of multiple copies of few-photon entangled states and combining them together with fusion operations \cite{Browne2005, Duan2005, Kieling2007}. This principle underlies the fusion-based quantum computation (FBQC) \cite{Bartolucci2023} framework, which maps the algorithm onto the sequence of reconfigurable two-qubit fusion measurements applied to a set of few-qubit entangled resource states. The entangled states, for example, Bell-states, are also consumed within a fusion operation circuit in order to increase the success probability \cite{Bartolucci2021}. 

Designing a linear optical circuit for generating entanglement is a challenging task, and optimal circuits have yet to be discovered. The heralded Bell state can be generated in the eight-mode \cite{Browne2005, Joo2007} and the six-mode \cite{Carolan2015, Stanisic2017, Gubarev2020} interferometers using four single photons with success probabilities of 1/16 and 2/27, respectively. Recently, a more compact five-mode scheme with an increased success probability of 1/9 was proposed \cite{Fldzhyan2021} and further extended to the entanglement-programmable layout \cite{Fldzhyan2023}. 

Early experiments used spontaneous parametric down-conversion sources and bulk optics for Bell states \cite{Zhang2008} and 4-GHZ states \cite{Zhang16} generation. Subsequently, the potential of integrated photonics was leveraged to implement heralded Bell state generation with a 2/27 success probability using a six-mode glass photonic chip \cite{Carolan2015}. Recent advances in the semiconductor quantum dot (QD) photon source fabrication with high brightness and indistinguishability \cite{Senellart2017, Arakawa2020, Tomm2021} enabled the generation of entangled states with much higher efficiency. The heralded generation of 3-GHZ states was demonstrated using six demultiplexed single photons from a single QD and both photonic chips and bulk optics \cite{Chen2024, Maring2024, Cao2024}. Larger 4-GHZ states were demonstrated using only four photons \cite{Li2020, Pont2024} and thus without heralding.

In this work, we present an experimental platform that enable performing small-scale experiments on linear-optical quantum computing targeted on the MBQC, FBQC and others alike models of quantum computing. We showcase the platform’s capability in producing heralded arbitrary two-qubit dual-rail encoded states~\cite{Fldzhyan2023}, a crucial building block for large-scale photonic quantum computers. In the experiment we have used a custom programmable 8-channel integrated photonic chip using femtosecond laser writing (FLW) and employed a high-quality QD-based single-photon source coupled with a low-loss demultiplexer to generate multiple photons simultaneously. We discuss the improvements to be made, in order to scale the platform further for more complex experiments.

\section{\label{sec:Methods}Experiment}

\subsection{\label{sec:level2}Linear optical scheme}

\begin{figure}[h!]
\includegraphics[width=1.0\columnwidth]{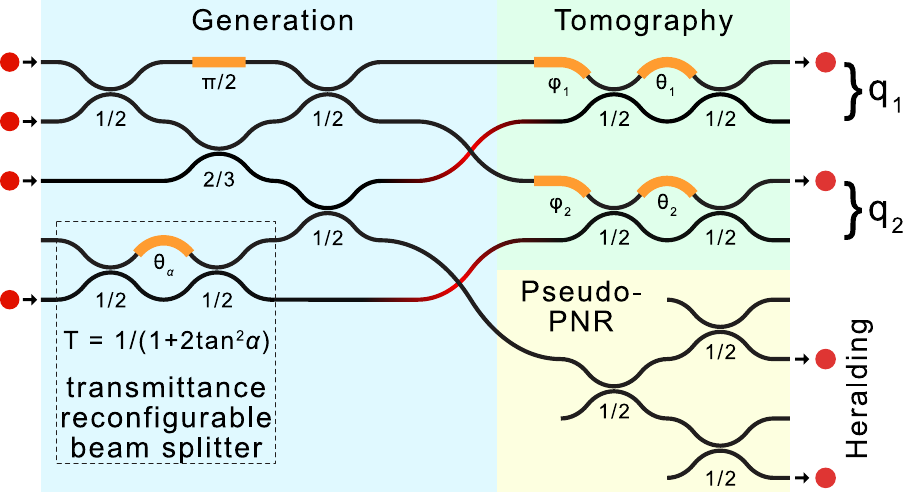}
\caption{\label{fig:1} The linear optical scheme for heralded two-qubit state generation (blue background color). Under the beam splitters and phase shifters, the transmittances and phases are indicated, respectively. The input MZI functions as a reconfigurable beam splitter $T(\alpha)$, which controls the entanglement of the generated state. The red lines represent the in-depth crossing waveguides using the 3D capabilities of the FLW. Two output MZIs are employed to perform QST (green background color). A cascade of three beam splitters at the output serves as a pseudo-PNR (yellow background color). Four single photons are injected into the input modes. The entangled two-qubit state in dual-rail encoding is heralded in the first four output modes by detecting a total of two photons in the last four output modes.}
\end{figure}

We utilized a linear optical programmable two-qubit state generation circuit, which was recently reported in \cite{Fldzhyan2023} (see Fig.~\ref{fig:1}). This scheme is capable of generating entangled two-qubit states with predefined entanglement degree controlled by the linear optical circuit, conveniently programmed by one phase shifter (PS) element. The injection of four single photons in the Fock state $| \phi_{in} \rangle = | 11101 \rangle$ generates a dual-rail-encoded two-qubit state in the first four output modes:

\begin{equation}
|\Phi(\alpha) \rangle = \cos \alpha |00 \rangle + \sin \alpha |11 \rangle,
\label{dual_rail_bell}
\end{equation}

where $\alpha$ is a parameter that controls the degree of entanglement. The parameter $\alpha$ is set via inner PS $\theta_\alpha$ of the input Mach–Zehnder interferometer (MZI), which effectively acts as a reconfigurable beam splitter (BS) with transmittance $T(\alpha) = 1/(1 + 2\tan ^2 \alpha)$. Two MZIs are applied to the output modes to perform single-qubit projection measurements for QST. The original work \cite{Fldzhyan2023} prescribed heralding the generated state by detecting two photons in the last optical mode. In this case, the success probability for state generation is given by the expression:

\begin{equation}
p(\alpha) = \frac{1}{6(1 + \sin^2\alpha)}, \quad (0 \leq \alpha \leq \pi/4).
\label{eq_2}
\end{equation}

We implement photon-number-resolving (PNR) photodetection by splitting the output mode using a cascade of three BSs and adding a single-photon detector at each of the four outputs. It can be calculated that, with a probability of 1/4 this system will remain unable to detect two photons. Consequently, the success probability for state generation from Eq.~\ref{eq_2} is multiplied by the measurement probability of 3/4.

\subsection{\label{sec:level2}Experimental setup}

\begin{figure}[ht!]
\includegraphics[width=1.0\columnwidth]{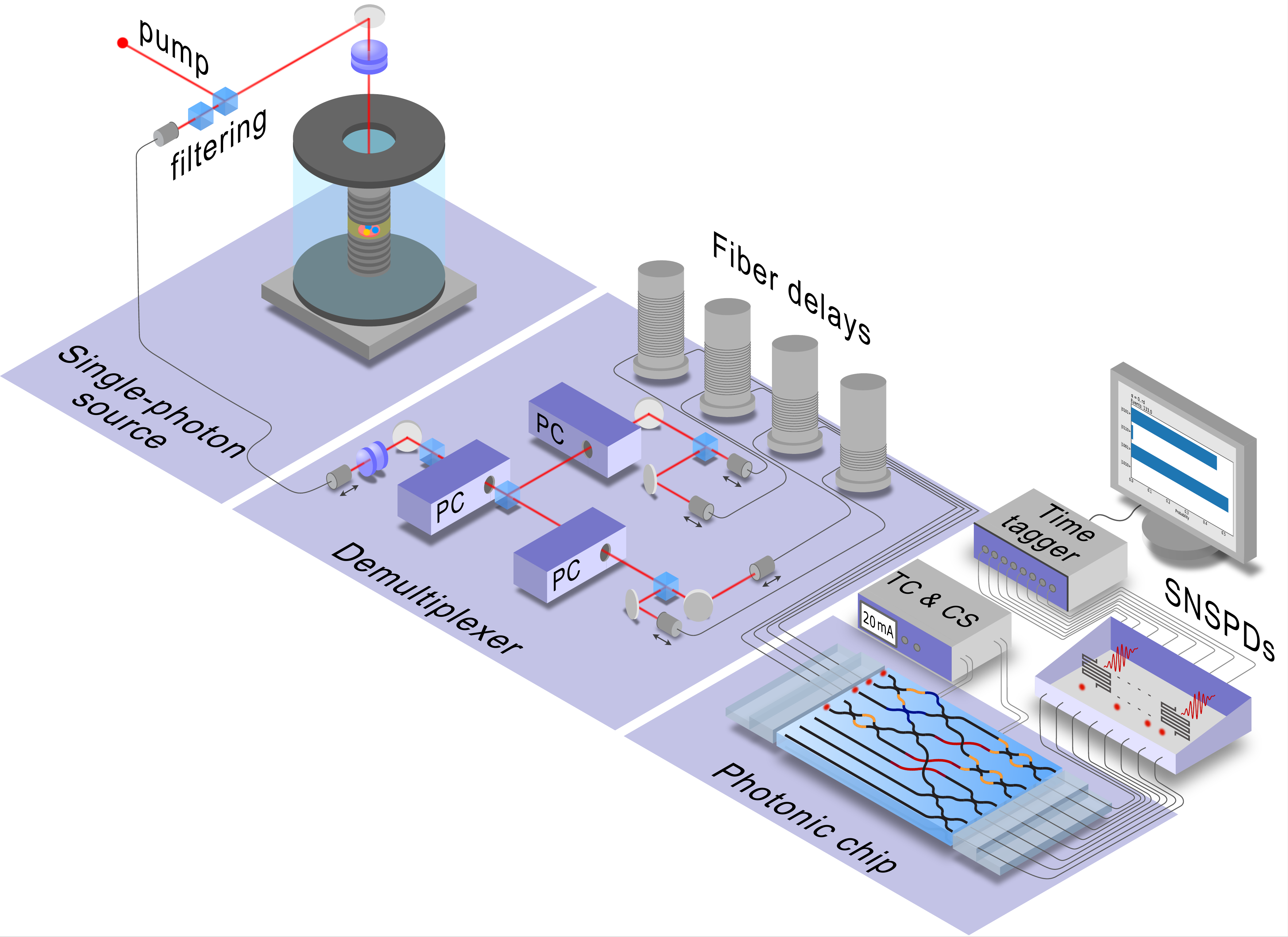}
\caption{\label{fig:2} Scheme of the experimental setup. A single photon source based on InAs/GaAs QD placed inside an optical cryostat emitted photons at a central wavelength of 920~nm. A demultiplexer in a tree configuration consisting of three Pockels cells (PCs) and fiber delays prepared four single photons, which were synchronized and injected into the 8-channel photonic chip in the Fock state $| \phi_{in} \rangle = | 11101000 \rangle$. The photonic chip was thermally stabilized using a temperature controller (TC) and programmed with a constant current source (CS). The output photons were detected using superconducting nanowire single-photon detectors (SNSPDs) and the detection signal was processed by a time tagger.}
\end{figure}

The experimental setup is depicted in Fig.~\ref{fig:2}. The quantum states were generated from four synchronized single photons, emitted from a single spatially-demultiplexed photon source. The photon source is an InAs/GaAs QD centered in a micropillar resonator \cite{RAKHLIN2023}. The sample is cooled down to 6K inside an optical cryostat. Spectrally filtered laser pulses generated by a picosecond fiber-laser source (Scoltech) with a repetition rate of 320 MHz (with an all-fiber frequency quadrupler \cite{Broome11}) resonantly pumped the trion transition in the QD in a cross-polarization configuration. Single photons at a central wavelength of 919 nm were coupled to a single-mode fiber with an efficiency of 11.4\%. The purity of single photons was estimated using a conventional Hanbury Brown-Twiss type measurement \cite{HANBURYBROWN1956}, which yielded  $g^{(2)}(0) = 0.048$.

The single-photon flux from the QD source was split into four independent channels with a demultiplexer consisting of three Pockels cells (PCs) (Leysop) and three polarizing beam splitters (PBSs) \cite{RAKHLIN2023}. The FPGA-based board receives a clock signal from the pump laser and outputs control pulses to PC high-voltage drivers (BME Bergmann) at the 10 MHz repetition rate.
The reflected photons are coupled to polarization maintaining (PM) single-mode fibers using adjustable collimators (Thorlabs, FiberPort) and passed through coiled fiber-optic delay lines of precisely tailored length in order to synchronize the photons at each output of the demultiplexer. The indistinguishability of single photons from different demultiplexer channels was determined using a Hong-Ou-Mandel (HOM) interferometer \cite{HOM87}. An average HOM interference visibility value of $0.97$ was obtained. The measured four-photon generation rate is 220 Hz with an average channel transmission efficiency of $65\%$. The measured $g^{(2)}(\tau)$ plots and the HOM interferences estimation algorithm are provided in Appendix~\ref{app:PhotonSource}. 

The optical scheme was fabricated using FLW technology, which enables the creation of low-loss photonic chips in fused silica glass \cite{skryabin2024}. The 8-channel photonic chip consists of 13 directional couplers (DCs), 3 crossers and 6 thermo-optic PSs, as shown in Fig.~\ref{fig:2}. The input and output channels of the chip were butt-coupled to V-groove PM fiber arrays mounted on six-axis positioners (Luminos). Thermo-optic PSs were created on the top surface by patterning resistive 30 $\mu$m wide and 4~mm long microheaters atop of the waveguides in a deposited nichrome layer via laser ablation process, which was performed in the same fabrication setup. The detailed description of the chip fabrication process is given in Appendix~\ref{app:Chip_fabrication}.
The PSs were calibrated and used to set the required phase $\theta_{\pi/2} = \pi/2$ and the phase $\theta_\alpha$ to establish the parameter $\alpha$, which controls the entanglement of the generated state, as well as the set of phases ($\phi_1, \theta_1; \phi_2, \theta_2$) in the output MZIs performing projective measurements. Also, the reflectivities of all DCs and the relative efficiencies of the output channels were obtained from the calibration data using an optimization algorithm. As a result, we reconstructed a complete numerical model of the photonic chip. A detailed description of the PSs calibration process is provided in Appendix~\ref{app:Chip_calibration}. To generate a heralded two-qubit state, four single photons were injected into input channels No. 1, 2, 3, 5 of the chip. The output photons from all eight channels were detected using commercially available superconducting nanowire single-photon detectors (SNSPDs, Scontel) and preprocessed using a time tagger (Swabian Instruments).

\section{\label{sec:level1}Results and Discussion}

\begin{figure}[ht!]
\includegraphics[width=1.0\columnwidth]{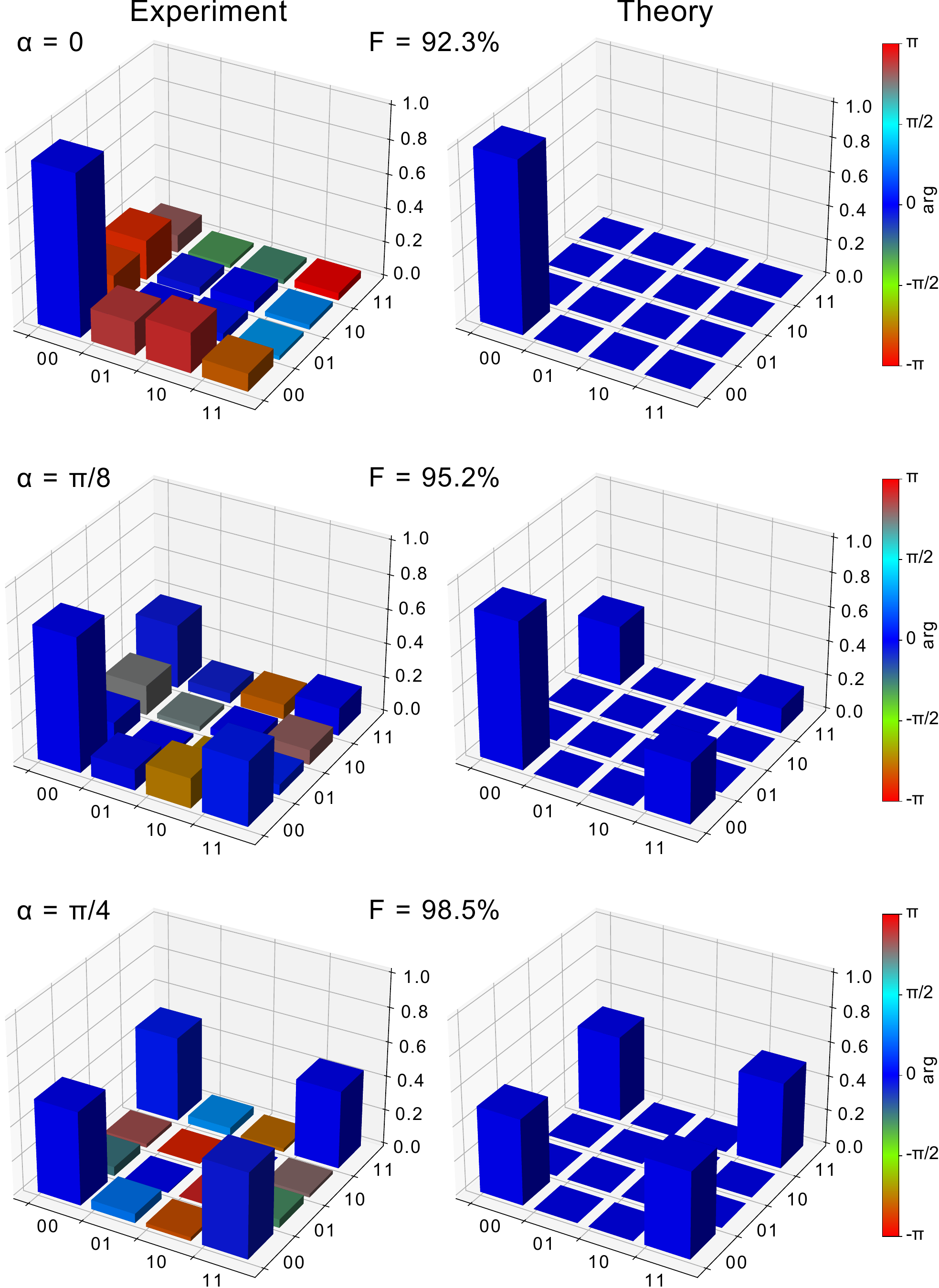}
\caption{\label{fig:3} Reconstructed density matrices of the generated states (left) for three values of the parameter $\alpha~=~0, \pi/8$ and $\pi/4$ and its ideal theoretical counterparts (right). The corresponding fidelities between them are shown above.}
\end{figure}

To demonstrate the capability of the scheme to generate two-qubit states with adjustable degree of entanglement, we chose three values of the parameter $\alpha~=~0, \pi/8$ and $\pi/4$,  corresponding to the factorized state $|00 \rangle$, the partially entangled state and the maximally entangled Bell states $(|00 \rangle + |11 \rangle)/\sqrt{2}$.  The MZIs placed at the output of the scheme enable us to implement projective measurements of the preparated states. We performed the measurements in six basis configurations by setting proper PS values in the MZIs and, based on the measurement data, reconstructed the state density matrices via QST. Imperfection of the photonic chip DCs reflectivities leads to state preparation and measurement (SPAM) errors. These errors are corrected by analyzing the numerical model of the chip, which helps to adjust the phase shifts in the state generation part, as well as to grasp information about the projectors, which are implemented in the tomography part. Details about the experiment and QST procedure are provided in Appendix~\ref{app:exp}. The reconstructed density matrices of the generated states for three values of the parameter $\alpha~=~0, \pi/8$ and $\pi/4$ and its ideal theoretical counterparts are shown in Fig.~\ref{fig:3}. The obtained fidelities between them are shown in Table~\ref{tab:table1} (middle column), where $F$ was defined as:

\begin{equation}
F(\rho_e, \rho_t) = \left(tr \sqrt{\sqrt{\rho_e}\rho_t \sqrt{\rho_e}}\right)^2.
\label{eq_4}
\end{equation}

\begin{table}[h!]
\caption{\label{tab:table1} Fidelity between experimentally obtained density matrices for different parameters $\alpha$ and its ideal theoretical and simulation (imperfections taken into account) counterparts.}
\begin{ruledtabular}
\begin{tabular}{ccc}
\makecell{Parameter $\alpha$} & \makecell{Fidelity \\ with $\rho_{th}$} & \makecell{Fidelity \\ with $\rho_{sim}$}\\
\hline
0 & 92.3 \% & 98.5 \%\\
$\pi/8$ & 95.2 \% & 98.8 \%\\
$\pi/4$ & 98.5 \% & 99.2 \%\\
\end{tabular}
\end{ruledtabular}
\end{table}

Besides the chip defects, the quality of the generated states is affected by single photon source imperfections. To take their contribution into account, we performed simulation using the Perceval software~\cite{heurtel2023perceval} with the experimentally determined probability of mutliphoton states at the output of the source and the partial distinguishability of photons. This helped us to obtain a much better agreement between the numerical predictions and the experimental results (see Table~\ref{tab:table1}, right column). The density matrices obtained from the numerical simulation are provided in Appendix~\ref{app:simul}.

To assess whether our reconstructed state is entangled or not, we use the Peres-Horodecki criterion \cite{peres1996separability, horodecki2006separability}, which is a necessary and sufficient criterion for a two-qubit system to be entangled. For instance, for the state in Fig. \ref{fig:3}(c) the negativity equals $0.495$, while for ideal Bell states this value is $1/2$, which is a proof that the reconstructed state is strongly entangled. 

\begin{figure}[ht!]
\includegraphics[width=1.0\columnwidth]{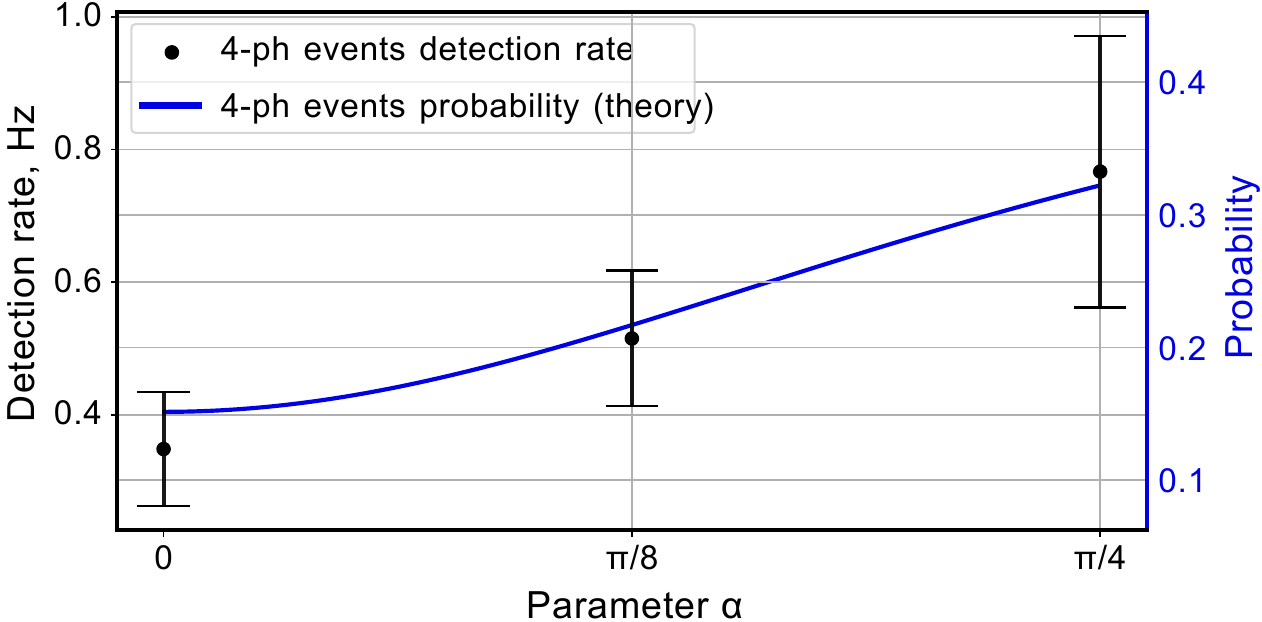}
\caption{\label{fig:4} The average values of the detection rate of four-photon events at different $\alpha$, obtained experimentally (black dots) and theoretical values of their fraction from all four-photon events (blue line).}
\end{figure}

During the experiment, we monitored the rate of four-photon events in all configurations possible. It was varied with the parameter $\alpha$ (see Fig.~\ref{fig:4}), since the generation rate depends on the entanglement degree of the target state \cite{Fldzhyan2023}. Since we detected only four separate photon events, we plotted a theoretical curve for the fraction of such four-photon events out of all four-photon events. The ratio between detected four separate photon events and this theoretical curve yields the detection rate of all four-photon events $f_{4-ph} = 2.36$~Hz. The efficiency of the entire setup $p$ was estimated by the equation $f_{4-ph} = f_{pump}/4 \cdot p^4$. The obtained value $p = 1.45 \%$ includes the efficiencies of the QD fiber coupling, demultiplexer, photonic chip including fiber array coupling, and SNSPDs detection. The estimated values are shown in Table~\ref{tab:table2} (middle column). To evaluate the heralding efficiency, we divided the successful events by the sum of all events where two photons were registered in the heralding channels and obtained value of $8.3 \pm 3.2 \times 10^{-5}$. 

Both state generation and heralding efficiency can be significantly increased by further optimizing each part of the experimental setup (see Table~\ref{tab:table2}, right column). The key element to augment is the single-photon source efficiency. Recently, a design of the QD with a fiber coupling efficiency above 50 \% was demonstrated \cite{Tomm2021}. Another path is to increase the transmission though a photonic chip. A reconfigurable circuit with nearly 60 \% ($\sim$2 dB) transmission was already used for a multiphoton experiment \cite{Pont2022}, and interferometers with total insertion loss below 20 \% ($\sim$1 dB) were demonstrated on their own \cite{Tan22, skryabin2024}. As for the rest, the demultiplexer requires high-quality bulk-optics elements and fiber-to-fiber couplers, and the SNSPDs with efficiency near 90 \% are already commercially available, while the possibility of achieving 99.5\% efficiency has been previously reported \cite{Chang2021}. Further, a fully integrated architecture can improve the efficiency of the entire setup which shows promise to bring photonic quantum computing within the feasible region of state-of-the-art technology \cite{Alexander2024}.

\begin{table}
\caption{\label{tab:table2} Performance and decomposed efficiency of the experimental setup at the current level and with possible improvements.}
\begin{ruledtabular}
\begin{tabular}{ccc}
\makecell{Parameter} & \makecell{This Work} & \makecell{Possible \\ improvement}\\
\hline
QD fiber coupling efficiency & 11.4 \% & 50 \% (Ref.\cite{Tomm2021}) \\
Demultiplexing efficiency & 50 \% & 90 \% \\
Photonic chip efficiency & 36.3 \% & 80 \% (Ref.\cite{Tan22, skryabin2024}) \\
Detection efficiency & 70 \% & 90 \% \\
4-ph. detection rate (w./o. chip) & 208 Hz & $\sim$2.2 MHz\footnote{Calculated} \\
4-ph. detection rate (w. chip) & 2.36 Hz & $\sim$1.45 MHz$^{\text{a}}$ \\

\end{tabular}
\end{ruledtabular}
\end{table}

The Bell-state generation scheme considered in this article \cite{Fldzhyan2023} is most effective with a success probability 1/9 if no feed-forward correction is used. Further, the success probability can be improved to 1/4 by using distillation and to 1/2 (2/3) by using bleeding approaches \cite{Bartolucci2021}.

\section{\label{sec:level1}Conclusion}

In conclusion, we have experimentally implemented a scheme for generating the heralded two-qubit state with arbitrary entanglement using a linear-optical platform. Our setup utilized laser written 8-channel photonic chip and a demultiplexed QD-based single-photon source. We demonstrate the two-qubit states generation from a separable state to a maximally entangled Bell state with a fidelity of 98.5 \%. Non-maximally entangled states are also of interest for quantum teleportation \cite{Modlawska2008, PATHAK2011}. We believe that the applied here approaches truly advance the development of resource states for MBQC and FBQC \cite{Bartolucci2023}.

\begin{acknowledgments}

The work was supported by Russian Science Foundation grant 22-12-00353 (https://rscf.ru/en/project/22-12-00353/) in part of Bell state measurement and interpretation of the results. The work was also supported by Rosatom in the framework of the Roadmap for Quantum computing (Contract No. 868-1.3-15/15-2021 dated October 5, 2021) in parts of Quantum dot fabrication (Contract No. R2152 dated November 19, 2021) and demultiplexed single photon source development, photonic chip fabrication and calibration (Contract No.P2154 dated November 24, 2021). S.P.K. and M.Yu.S. acknowledges support by Ministry of Science and Higher Education of the Russian Federation and South Ural State University (agreement №075-15-2022-1116). Yu.A.B and S.A.F. are grateful to the Russian Foundation for the Advancement of Theoretical Physics and Mathematics (BASIS) (Projects №24-2-10-57-1 and №23-2-10-15-1).

\end{acknowledgments}

\section*{Author Declaration}

The authors have no conflicts to disclose.

\section*{Data Availability Statement}

The data that support the findings of this study are available from the corresponding author upon reasonable request.

\appendix

\section{\label{app:PhotonSource} Demultiplexed single photon source}

\begin{figure*}[t!]
\includegraphics[width=2.0\columnwidth]{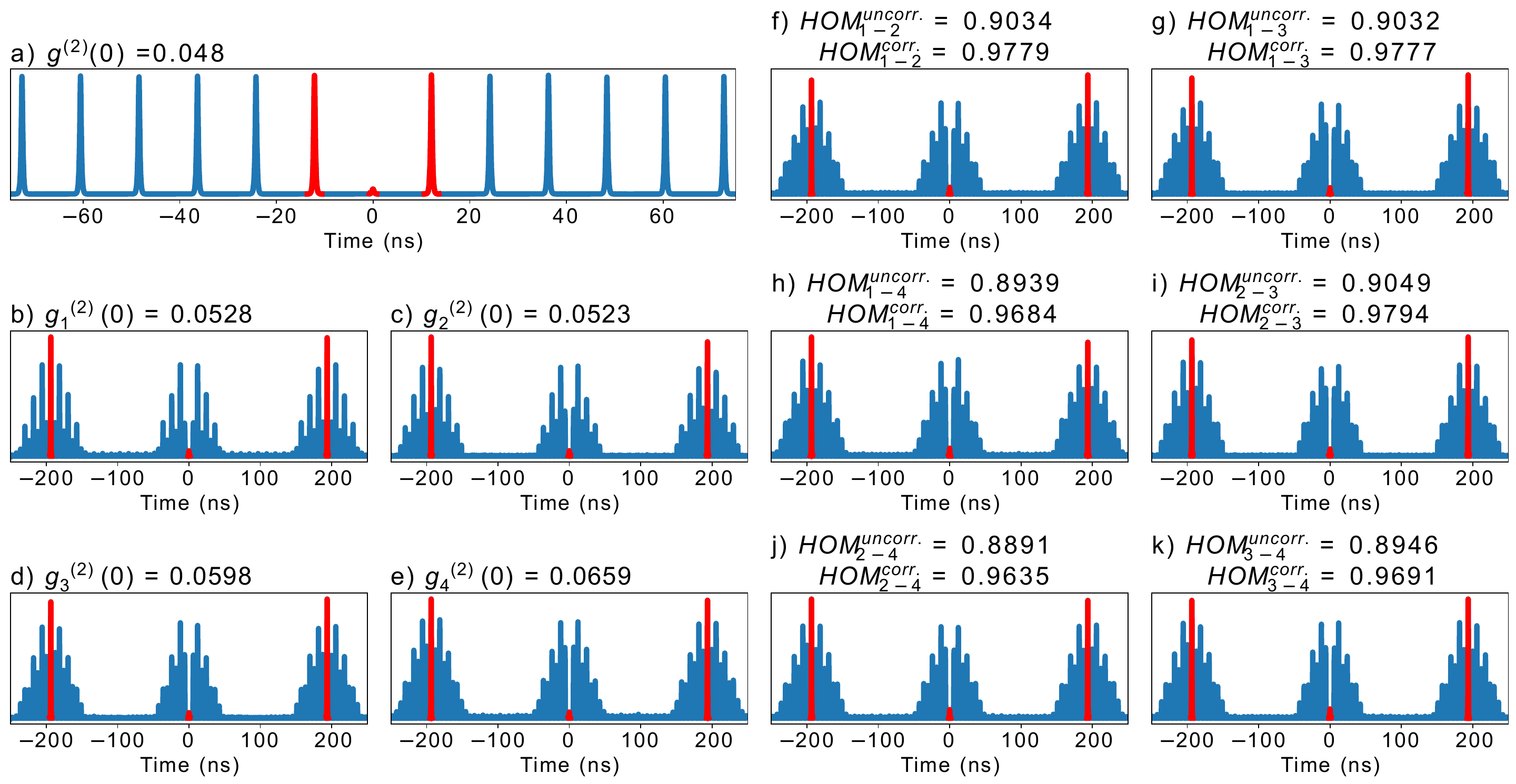}
\caption{\label{fig:A1} (a) $g^{(2)}(0)$ for photons from QD; (b)-(e) $g^{(2)}(0)$ for photons from demultiplexer channels and (f)-(k) HOM for pairs of demultiplexer channels measured using the DC in the photonic chip. Peaks used to determine numerical parameters are highlighted in red.}
\end{figure*}

The autocorrelation function of photons from QD $g^{(2)}(0)$ was measured using a fiber beam splitter and is shown in Figure~\ref{fig:A1}(a). After demultiplexing, the autocorrelation function of each channel of the demultiplexer $g^{(2)}_i(0)$ was measured using the lowest DC (designed to resolve the number of photons) of the photonic chip (Figure~\ref{fig:A1}(b)-(e)).

In both cases, the numerical values of the function were determined by:
\begin{equation}
g^{(2)}_i(0) = \frac{2A_c}{A_l+A_r},
\label{eq_g2}
\end{equation}
where $A_c$ is the area of the central peak, $A_l$ and $A_r$ are the areas of the neighboring left and right peaks, respectively.

The visibility of HOM interference for photons from the demultiplexer channels in pairs was also measured using the DC in the photonic chip with a reflectivity $R=0.4979$ (measured experimentally). The uncorrected $HOM^{uncorr}_{i-j}$ value was found using:
\begin{equation}
HOM^{uncorr.}_{i-j} = 1-\frac{2A_c}{A_l+A_r}.
\label{eq_HOM_uncorr}
\end{equation}

The obtained values must be corrected taking into account the visibility of classical interference $1-e = 0.95$ and the value of the autocorrelation function of the photon source $g^{(2)}(0)$. The corrected values of $HOM^{corr}_{i-j}$ were calculated according to:
\begin{multline}
HOM^{corr.}_{i-j} = \frac{1}{1-e^2}\Bigl(\frac{3g^{(2)}(0)}{2} +\frac{R^2+T^2}{2RT}\\ -\frac{R^2+T^2}{2RT}\frac{2A_c}{A_l+A_r}\Bigr).
\label{eq_HOM_corr}
\end{multline}

The final corrected HOM values for all pairs of channels are equal to $HOM^{corr.}_{1-2} = 0.9779$, $HOM^{corr.}_{1-3} = 0.9777$, $HOM^{corr.}_{1-4} = 0.9684$,  $HOM^{corr.}_{2-3} = 0.9794$, $HOM^{corr.}_{2-4} = 0.9635$ and $HOM^{corr.}_{3-4} = 0.9691$.

\section{\label{app:Chip_fabrication}Photonic chip fabrication and characterization}

\begin{figure}[ht!]
\includegraphics[width=0.75\columnwidth]{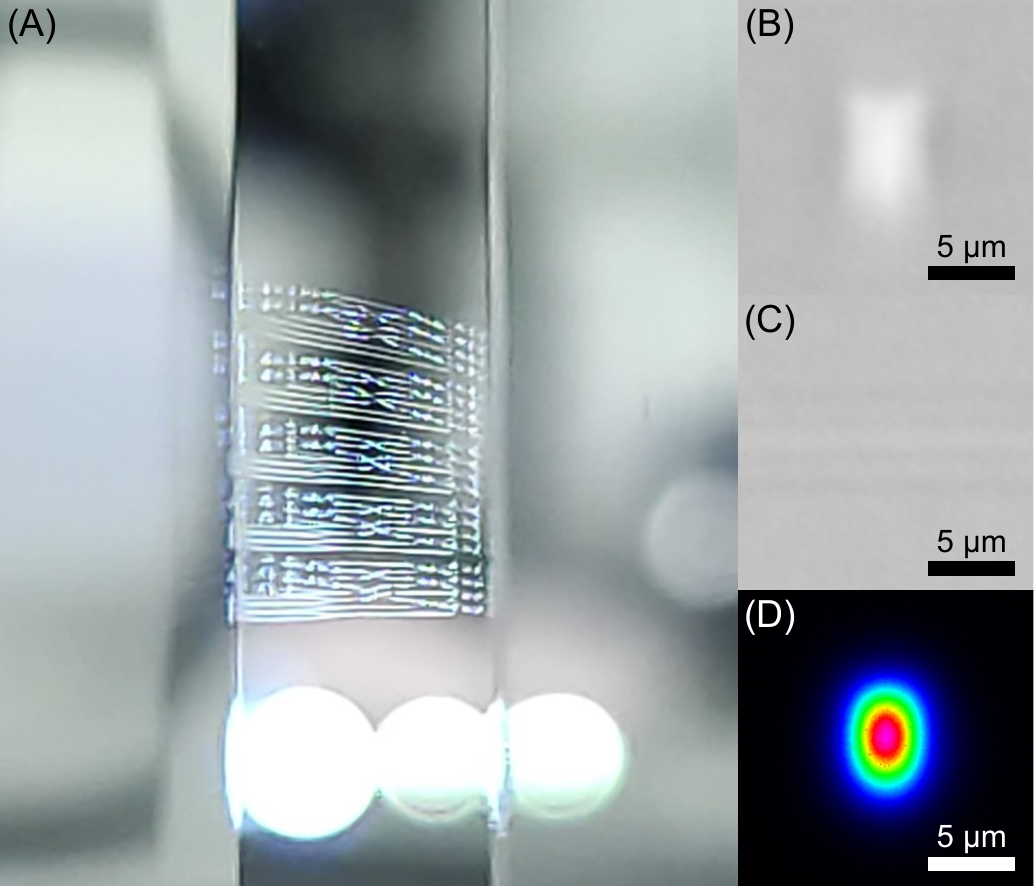}
\caption{\label{fig:A2} (a) Photograph of the photonic chip containing five waveguide structures. (b) Facet 
 and (c) top view micrographs of the waveguide, and (d) corresponding mode profile for V-polarization.}
\end{figure}

The photonic chip was fabricated by the FLW technology in fused silica sample (JGS1 glass) with lengths of 5~cm. Photograph of the photonic chip containing five waveguide structures is shown in Fig.~\ref{fig:A2}(a). The waveguides were written using power stabilized femtosecond pulses from a frequency doubled ytterbium fiber laser (Avesta Antaus) with the central wavelength of 515~nm, pulse duration of 270~fs, pulse energy of 47~nJ and the repetition rate of 1~MHz focused with an aspheric lens (NA = 0.55) 15 $\mu$m below the surface. A variable beam expander was used to increase the laser beam diameter in order to completely fill the entrance aperture of the focusing lens. The waveguides were written using the multiscan principle and consist of 21~scans with an offset of 0.2~$\mu$m. During each scan the sample was translated in one direction relative to the focal spot with a high-precision stage (AeroTech ANT) at a writing speed of 4~mm/s. Circular polarization of the laser beam was set by a quarter-wave plate $\lambda/4$. The waveguides facet and top view micrographs are shown in Fig.~\ref{fig:A1}(b) and (c), respectively. The waveguides support a single guided mode with mode field diameters 5.5~$\mu$m~$\times$~7.6~$\mu$m for V-polarization (see Fig.~\ref{fig:A1}d) and exhibit 0.07~dB/cm propagation loss and 0.8 dB coupling loss (with PM-780HP fiber) at 920~nm wavelength for V-polarization. The FLW setup scheme and more detailed information about the multiscan waveguides provided in \cite{skryabin2024}. Directional couplers consist of two waveguides with two s-bends, which bring cores together to the distances of 7.1~$\mu$m and 6.5~$\mu$m corresponding to 50:50 and 33:67 power splitting ratios, respectively. The s-bends are the combination of two circular arcs with radius of 60~mm. Crossers are spaced at a depth of 20~$\mu$m using the 3D capabilities of the FLW. The distance between the waveguides at the input/output facet and inside the circuit was set to match the fiber array pitch of 127~$\mu$m. The input and output facets of the sample were polished to optical quality after the writing process.

\begin{figure}[h!]
\includegraphics[width=1.0\columnwidth]{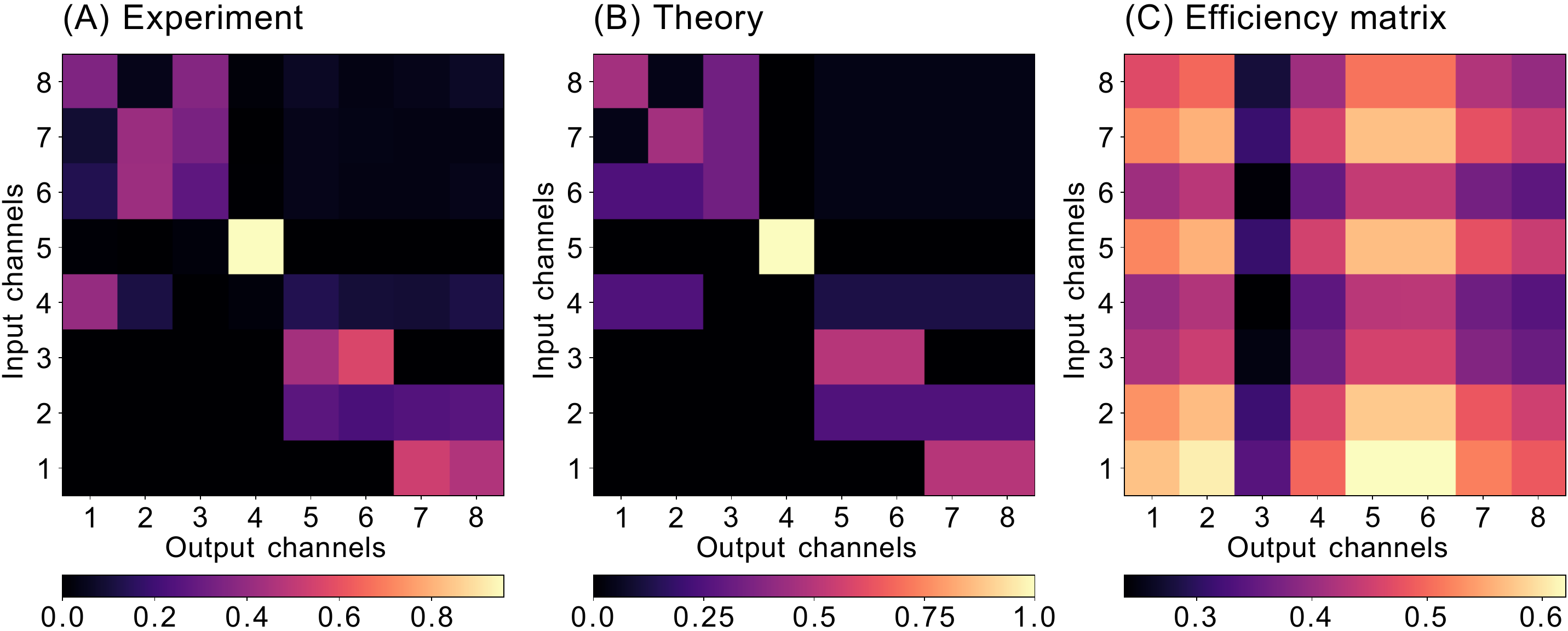}
\caption{\label{fig:A3} (a) Experimentally obtained doubly stochastic chip transfer matrix. (b) Theoretical chip transfer matrix. (c) Chip efficiency matrix.}
\end{figure}

All five structures inside photonic chip were experimentally characterized using radiation from a CW-laser. In order to obtain a structure transfer matrix the laser radiation was injected one by one into all input channels and collected from all output channels. The obtained matrix was reduced to a doubly stochastic form using the Sinkhorn-Knopp algorithm, which also produces the chip efficiency matrix. The obtained transfer matrix for the best structure and its theoretical ideal counterpart, as well as the efficiency matrix are shown in Fig.~\ref{fig:A3} (a), (b) and (c), respectively. Fidelity between these transfer matrices is F =  99.64\%, where fidelity was calculated according to:

\begin{equation}
F(M_{e}, M_{t}) = \frac{|Tr(M_{e}^{\dagger}M_{t})|^2}{Tr(M_{e}^{\dagger}M_{e})Tr(M_{t}^{\dagger}M_{t})}.
\label{Fid_eq}
\end{equation}

Variations in the chip efficiency matrix are due to differences in fiber coupling efficiencies for different input and output channels. The efficiencies ranged from 0.26 to 0.62, corresponding to losses from 5.85 to 2.08 dB. A significant drop in chip efficiency is noticeable in the output channel 3, which is associated with a defect on the chip output facet.

\section{\label{app:Chip_calibration}Phase shifters calibration and numerical model of the photonic chip}

\begin{figure}[h!]
\includegraphics[width=1.0\columnwidth]{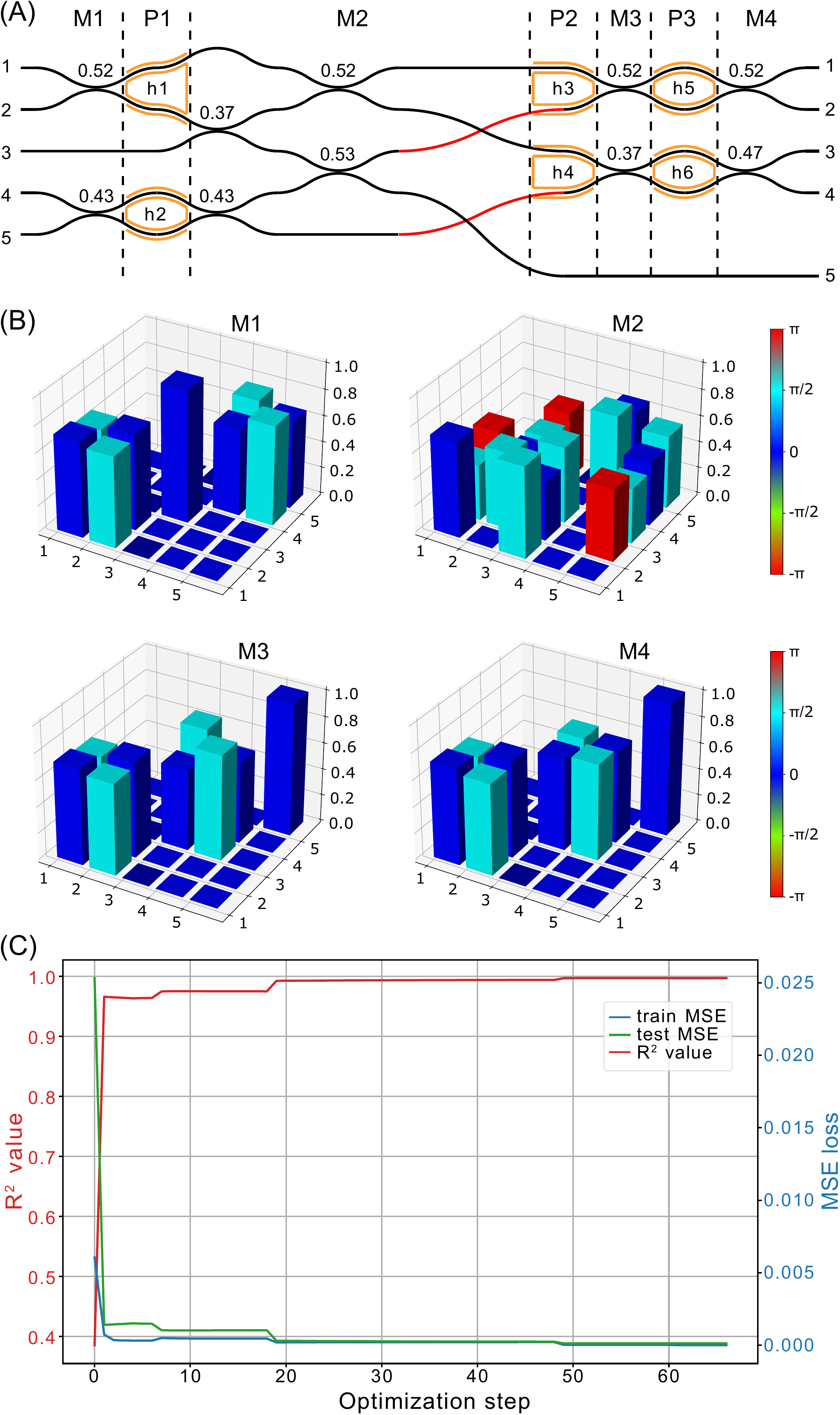}
\caption{\label{fig:A4} (a) The 5-channel photonic chip scheme divided into layers according to the numerical model. The obtained reflectivities $R$ are shown above corresponding DCs. (b) The obtained unitary matrices of mixing layers. (c) The variation in MSE loss on train and test datasets and $R^2$ value during the optimization process.}
\end{figure}

To calibrate thermo-optic PSs we injected single photons one by one into each input channel and measured photocounts from all output channels as a function of the applied current to particular PS. Thus, for each of the six PSs and for each of the five input channels, five experimental dependences of the photocounts on the current in the range from 0 to to 19,5 mA were obtained. All calibration data was then split into samples of the form $(port, \vec{x}, \vec{p})$ and used to train the chip model with machine learning algorithm, where $port$ is a number of input channel where photons were injected, $\vec{x}$ - array of six currents on PSs, $\vec{p}$ - normalized output photocounts distribution. 

The photonic chip numerical model consisted of phase layers $P_1, P_2, P_3$ and mixing layers $M_1, M_2, M_3, M_4$ (see Fig.~\ref{fig:A4}a). The unitary matrix of each phase layer had a known diagonal form and depended on two phases - controlled parameters. For example, 
\begin{equation}
    P_1(\phi_1,\phi_2) = \begin{pmatrix}
e^{i\phi_1} & 0 & 0 & 0 & 0  \\
0 &1 &0 &0 & 0\\
0 &0 &1 &0 & 0 \\
0 &0 &0&e^{i\phi_2}& 0 \\
0 &0 &0 &0 & 1 
\end{pmatrix}.
\end{equation}
In order to take into account crosstalk between PSs in each phase layer, the phases on the PSs $\vec{\phi}$ were related to the corresponding current strengths $\vec{x}$ as follows:
\begin{equation}
    \begin{pmatrix}
        \phi_1\\ \phi_2 
    \end{pmatrix}= \begin{pmatrix}
        \phi_{01}\\ \phi_{02}
    \end{pmatrix}+\begin{pmatrix}
        \alpha_{11}&\alpha_{12}\\
        \alpha_{21}&\alpha_{22}
       
    \end{pmatrix} \begin{pmatrix}
        x_1^2\\ x_2^2 
    \end{pmatrix}, 
\end{equation}
where $\vec{\phi_0}$ - the initial phase vector and $A = \begin{pmatrix}
        \alpha_{11}&\alpha_{12}\\
        \alpha_{21}&\alpha_{22}
       
    \end{pmatrix}$ - crosstalk matrix, whose elements were also trainable parameters.

The unitary matrix of the mixing layers was determined by the reflectivity $R$ of the DCs of which it consisted as:

\begin{equation}
    U_{DC}(R)=\begin{pmatrix}
        \sqrt{R}&i\sqrt{1-R}\\
        i\sqrt{1-R}&\sqrt{R}     
    \end{pmatrix}. 
\end{equation}

For example, the matrix of the mixing layer $M_2$ had the following form:
\begin{equation}
    M_2(R_1, R_2, R_3, R_4) = SWAP\cdot M_2''(R_3, R_4)\cdot M_2'(R_1,R_2),
\end{equation} 
where:
\begin{equation}
SWAP=\begin{pmatrix}
        1&0&0&0&0\\
        0&0&1&0&0\\
        0&1&0&0&0\\
        0&0&0&0&1\\
        0&0&0&1&0\\
    \end{pmatrix},
\end{equation}
\begin{equation}
M_2'(R_1,R_2)=\begin{pmatrix}
        1&0&0&0&0\\
        0&\sqrt{R_1}&i\sqrt{1-R_1}&0&0\\
        0&i\sqrt{1-R_1}&\sqrt{R_1}&0&0\\
        0&0&0&\sqrt{R_2}&i\sqrt{1-R_2}\\
        0&0&0&i\sqrt{1-R_2}&\sqrt{R_2}\\
    \end{pmatrix},
\end{equation}
\begin{equation}
M_2''(R_3,R_4)=\begin{pmatrix}
        \sqrt{R_3}&i\sqrt{1-R_3}&0&0&0\\
        i\sqrt{1-R_3}&\sqrt{R_3}&0&0&0\\
        0&0&\sqrt{R_4}&i\sqrt{1-R_4}&0\\
        0&0&i\sqrt{1-R_4}&\sqrt{R_4}&0\\
        0&0&0&0&1\\
    \end{pmatrix}.
\end{equation}

The unitary 5x5 matrix of the photonic chip was obtained by multiplying the layer matrices:
\begin{equation}
    U(\vec{x}, \vec{R}, A, \vec{\phi_0}) = M_4 \cdot P_3 \cdot M_3 \cdot P_2 \cdot M_2 \cdot P_1 \cdot M_1.
\end{equation} 

Since the measured photocounts distributions from output channel were affected by fiber coupling and photon detection efficiencies, we introduce a single vector $\vec{T}_{out}$ of the output channels efficiency. The final vector was renormalized to sum to 1. Thus, the total number of trainable parameters of the photonic chip model was 33: 12 elements of the three crosstalk matrices $A_{1-3}$, 6 elements of initial phases $\vec{\phi_0}$, 10 elements of the DCs reflectivity $\vec{R}$ and 5 elements of the output channels efficiencies $\vec{T}_{out}$.

All 1830 experimental data samples $(port, \vec{x}, \vec{p})$ were then split into train $n_{train}$ and test $n_{test}$ samples with a ratio 80/20. We used Adam optimizer of PyTorch package in Python to update model parameters $A$, $\vec{\phi_0}$, $\vec{R}$ and $\vec{T}_{out}$ through stochastic gradient-descent algorithm to minimize the mean squared error (MSE) between true labels $p$ of training dataset and model predictions $\hat{p}$ for output distributions. MSE is defined by the formula:
\begin{equation}
    MSE(p,\hat{p}) = \frac{1}{5}\sum_{i=1}^{5}(\hat{p}_i-p_i)^2.
\end{equation}
The initial values of the training parameters were chosen as follows: $\vec{\phi_0}^{(i)}=0$, $\vec{R}^{(i)}=0.5$, $\vec{T}_{out}^{(i)}=1$, all crosstalk matrices $A = \begin{pmatrix}
        2.14&0.53\\
        0.53&2.14
       
    \end{pmatrix}\cdot 10^{-2} \frac{rad}{mA^2}$.
After each step through the training dataset the $R^2$ value and MSE on the test dataset were evaluated. The coefficient of determination $R^2$ is defined by the formula:
\begin{equation}
    R^2(y,\hat{y}) = 1-\frac{\sum_{i=1}^{5N}(y_i-\hat{y}_i)^2}{\sum_{i=1}^{5N}(y_i-\overline{y}_i)^2},
\end{equation}
where $N$ is the test dataset size, $y$ is the vector $1 \times N$ composed of all elements $p_i$ of all test dataset vectors $\vec{p}$, $\hat{y}$ is the vector $1 \times N$ composed of all elements $\hat{p}_i$ of all model predictions $\hat{p}$ on test dataset, $\overline{y} = \frac{1}{N}\sum_{i=1}^{i=N}y_i$. 

As a result of the model training optimized values for the parameters $A$, $\vec{\phi_0}$, $\vec{R}$ and $\vec{T}_{out}$ were found. The obtained reflectivities $R$ of the DCs are shown above the corresponding DC in Fig.~\ref{fig:A4}a, and also as mixing layers unitary matrices (see Fig.~\ref{fig:A4}b). The optimized crosstalk matrices $A_{1-3}$ presented in $\frac{rad}{mA^2}$ and initial phases of the PSs, as well as the vector of the output channels efficiencies $\vec{T}_{out}$ were obtained as follows:

$A_1 = \begin{pmatrix}
        2.24&0.53\\
        0.40&2.03
    \end{pmatrix}\cdot 10^{-2}$;
$\vec{\phi_{01}} = \begin{pmatrix}
        -1.71\\
        0.21
    \end{pmatrix}$

$A_2 = \begin{pmatrix}
        1.71&0.82\\
        0.53&3.57
    \end{pmatrix}\cdot 10^{-2}$;
$\vec{\phi_{02}} = \begin{pmatrix}
        1.36\\
        -3.01
    \end{pmatrix}$

$A_3 = \begin{pmatrix}
        3.06&0.78\\
        0.93&2.62
    \end{pmatrix}\cdot 10^{-2}$;
$\vec{\phi_{03}} = \begin{pmatrix}
        -0.80\\
        0.17
    \end{pmatrix}$

$\vec{T_{out}} = \begin{pmatrix}
        0.69;\
        1.00;\
        0.59;\
        0.94;\
        0.76\
    \end{pmatrix}$,
where the maximum value of $\vec{T}_{out}$ was normalized to 1.

\begin{figure}[h!]
\includegraphics[width=1.0\columnwidth]{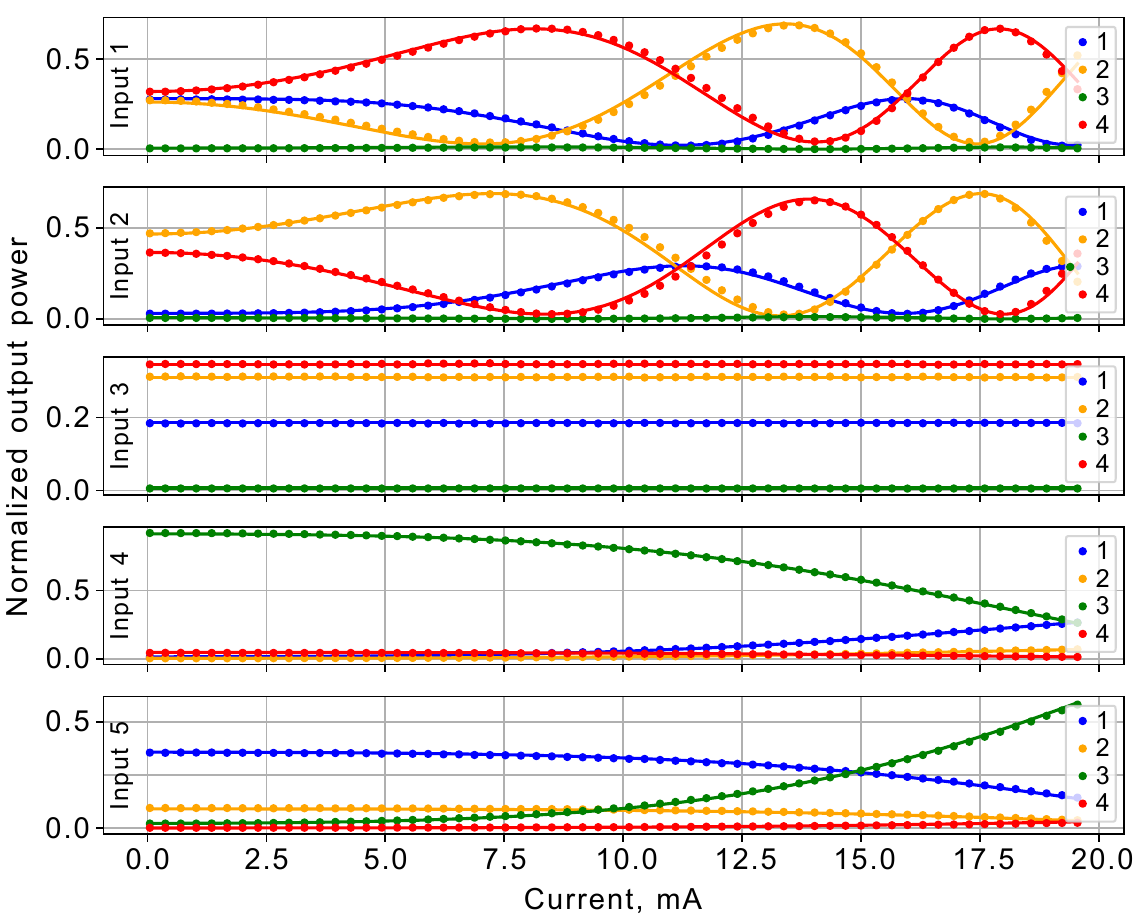}
\caption{\label{fig:A5} Calibration data for PS$_{1}$ approximated by the numerical model of the photonic chip.}
\end{figure}

The variation in MSE loss on train and test datasets and $R^2$ value on test dataset during optimization process are shown in the Fig.~\ref{fig:A4}c. The total $R^2$ value on the entire data was equal to 0,9982. An example of calibration data approximated using the numerical model of the photonic chip is shown in Fig.~\ref{fig:A5}, which shows a fairly good accuracy of the numerical model.

\section{Experiment and Quantum state tomography}
\label{app:exp}

In order to set the desired parameter $\alpha$, one need to calculate the required phase $\theta_{\alpha}$ by solving a system of equations:

\begin{eqnarray}
    U_{MZI}(\theta_{\alpha})= U_{DC}(R_2) \cdot P(\theta_{\alpha}) \cdot U_{DC}(R_1),
\\
    T(\alpha) = |U_{MZI10}(\theta_{\alpha})|^2 = 1/(1 + 2\tan ^2 \alpha).
\end{eqnarray}
For a set of parameters $\alpha~=~0, \pi/8$ and $\pi/4$ the corresponding phases are $\theta_\alpha~=~0,  -0.337\pi$ and $-0.608\pi$ for ideal DCs with reflectivities $R$ = 0.5. 

But unfortunately, in our chip the reflectivities of the DCs have deviations from ideal ones. This inevitably leads to SPAM errors. We decided to optimize state generation part by two PSs ${\theta_{\pi/2}, \theta_{\alpha}}$ to make the output state closer to expected. To do this, we use numerical model of the photonic chip, obtained during the PSs calibration process, and numerically simulate output state from generation part of the scheme. Those phases were found for all the three values of $\alpha = 0, \pi/8, \pi/4$ and equal to ${\theta_{\pi/2}, \theta_{\alpha}}~=~ {(0.44\pi, 0.03\pi), ( 0.57\pi, -0.27\pi), (0.54\pi, -0.56\pi)}$, respectively. 

The  multiphoton event counts for each output configuration were corrected to account for the efficiency of the output channels, which includes the chip-to-fiber array coupling and the SNSPDs detection efficiencies. 

A comparison of simulated fidelities between ideal states and those with non-optimized and optimized phases are shown in Table~\ref{tab:table3}. Even though for $\alpha = 0$ chip imperfections have not affected much the output state, it is clear that in case of non-ideal chip structure one can not obtain high-fidelity correspondence between output states and those expected theoretically for other values of $\alpha$ without proper choice of phases. The main thing that affects fidelity in case of non-optimized phases is relative phase factor between $|00\rangle$ and $|11\rangle$ terms, while the value of $\alpha$ is relatively close to expected. This explains why in case $\alpha = 0$ the fidelity is still high.

\begin{table}
\caption{\label{tab:table3} Simulated fidelity between ideal density matrices for different parameter $\alpha$ and its counterparts with optimized $\rho_{opt}$ and non-optimized $\rho_{nopt}$ phases.}
\begin{ruledtabular}
\begin{tabular}{ccc}
\makecell{Parameter $\alpha$} & \makecell{Fidelity \\ with $\rho_{nopt}$} & \makecell{Fidelity \\ with $\rho_{opt}$}\\
\hline
0 & 98.2 \% & 98.9 \%\\
$\pi/8$ & 87.5 \% & 99.6 \%\\
$\pi/4$ & 84.1 \% & 99.5 \%\\
\end{tabular}
\end{ruledtabular}
\end{table}

\begin{figure}[h!]
\includegraphics[width=1.0\columnwidth]{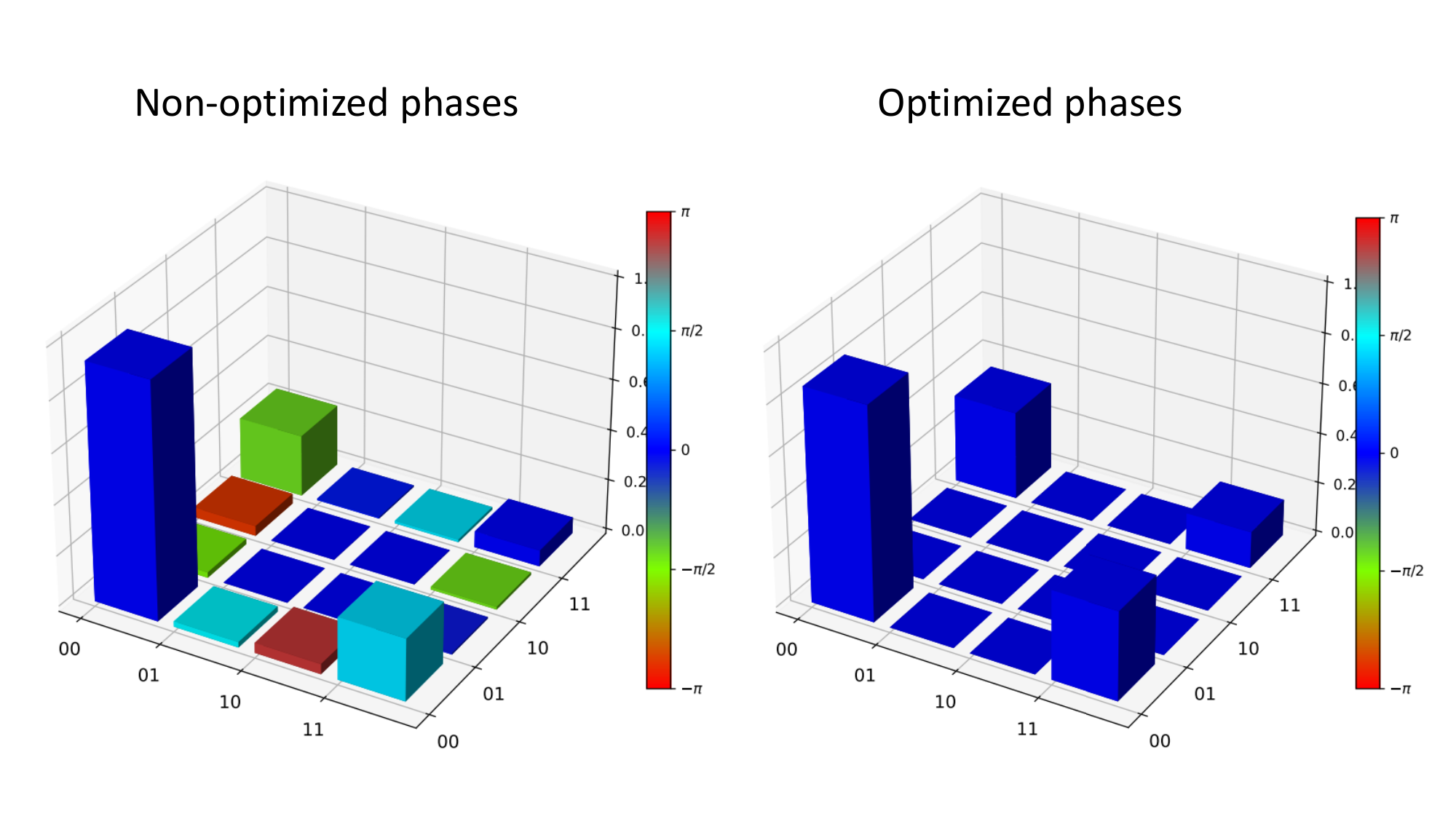}
\caption{\label{fig:A6} Comparison between output states in case of optimized phases (on the right image) and non-optimized (on the left image) for $\alpha = \pi/8$. The main difference between them is non-zero relative phase between $|00\rangle$ and $|11\rangle$ terms.}
\end{figure}

We implemented the procedure of QST to reconstruct the density matrix of logical state. Since the dimension of Hilbert space of two qubits equals 4, the reconstructed density matrix has only 15 independent parameters. We have measured our state in four different bases (combinations of different single-qubit bases) to obtain 16 independent values, characterising it. We have implemented maximum likelihood estimation \cite{baumgratz2013scalable} to get the density matrix of generated state.

Since the DCs in the output MZIs also are not ideal, basis in which we measure our state differ from expected in ideal case. For instance, measurement in h-basis, when  the output state of dual-rail qubit is projected on states $|01\rangle$ or $|10\rangle$ can only be set if both DCs comprising MZI have the same splitting ratio, which is not true in our case. To get reed of those measurement errors we calculate true projectors that we measure using numerical model of the photonic chip.

\section{\label{app:simul}Simulations with source imperfections}

\begin{figure}[h!]
\includegraphics[width=1.0\columnwidth]{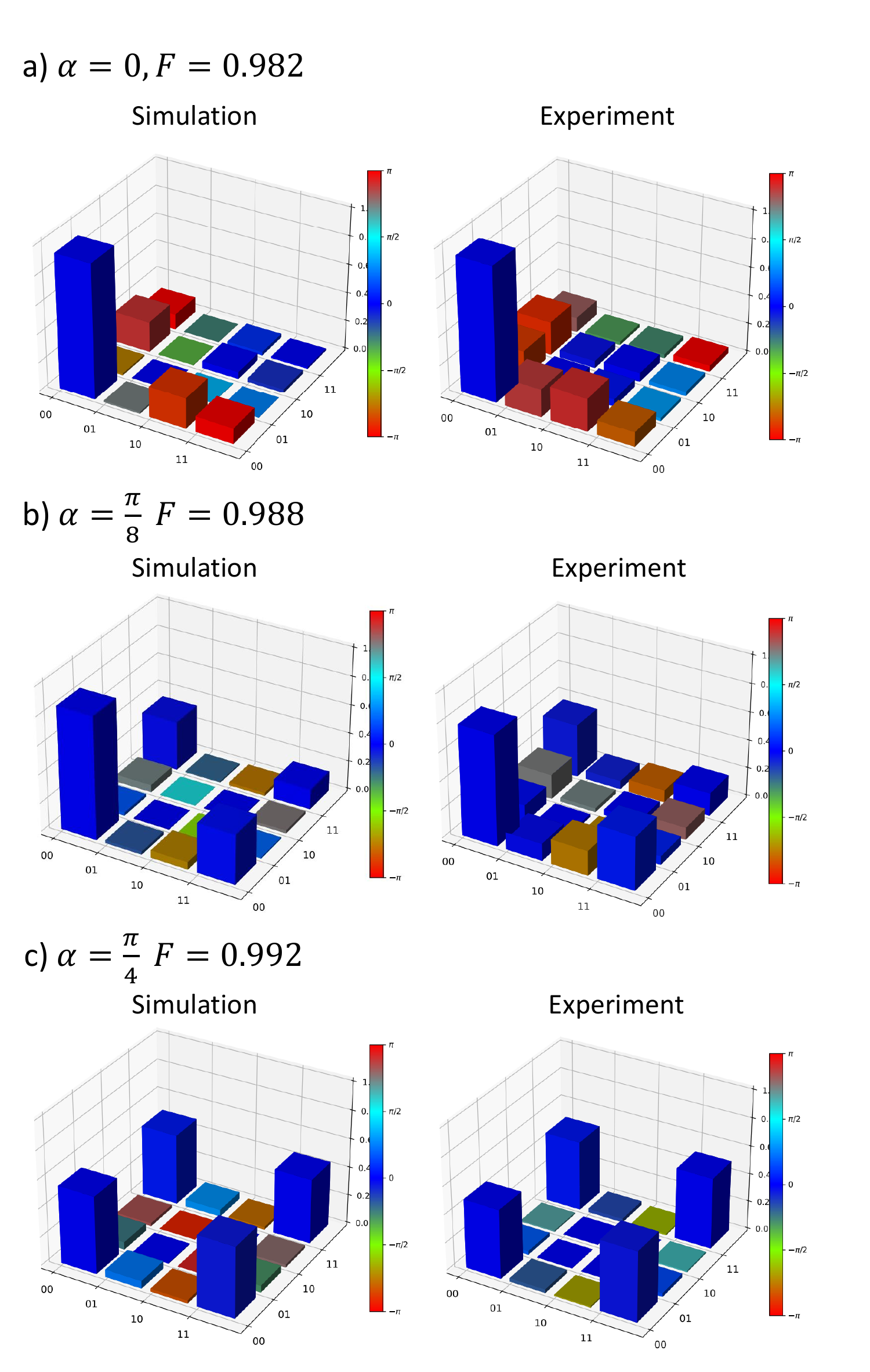}
\caption{\label{fig:A7} Comparison of density matrices for $\alpha = 0, \pi/8, \pi/4$ obtained by numerical simulation with all the source and chip imperfections and experimentally reconstructed.}
\end{figure}

The density matrices obtained by numerical simulation and experimentally reconstructed ones are shown in Fig.~\ref{fig:A7}

\section*{References}

\bibliography{aipsamp}

\end{document}